# Enhancing Volumetric Optical Chirality through 2D–3D Structural Design Evolution

Chia-Te Chang,[†] Xiaoyan Zhou,[†] Dmitrii Gromyko, John You En Chan, Lin Wu, Chia-Ming Yang,* Sejeong Kim,* Hongtao Wang,* and Joel K.W. Yang*

## Abstract:

Circular dichroism (CD) sensing plays a pivotal role in probing molecular chirality in biomedical sciences. However, engineering superchiral electromagnetic fields that can reliably amplify the faint signatures of chiral analytes remains profoundly challenging. Central to this difficulty is the need to balance two competing demands: maximizing the enhancement of chiral fields while maintaining a sufficiently large interaction volume for effective molecular interrogation. Here, we introduce a figure of merit (FOM) that captures the enhancement and spatial coverage of superchiral fields to benchmark different chiral-field configurations. We examine the effects of helix-geometry evolution on the FOM, including 2D to 3D chirality induction, winding-number escalation, helical-order enhancement, and transverse dilation. By tuning these structural degrees of freedom, the sensing volume can be enlarged without compromising the distribution and enhancement strength of fields. The optimized triple-strand helix markedly enhanced the analyte CD signal, yielding a FOM of $2.43 \times 10^{10}$ nm$^3$, which surpassed prior 2D and 3D configurations by over an order of magnitude. The proposed FOM exhibits a strong linear correlation ($R^2 = 0.9256$) with the analyte CD signal. Our findings provide a systematic design framework for 3D chiral structures and a robust metric for assessing their chiroptical sensing performance, particularly in scenarios involving clusters of randomly oriented small molecules or a large chiral molecule.

Chirality, a fundamental geometric property of an object that cannot be superimposed on their mirror images, is a fundamental characteristic of many molecules in nature. A pair of chiral molecules, consisting of an original structure and its mirrored one are called enantiomers. Although enantiomers share the same molecular formula but exhibit different chemical and biological properties.[1-3] For example, the different chirality of protein secondary structures can significantly influence their interaction with the human body,[4] where one chiral form may be benign while another can be harmful.[5, 6] Understanding molecular chirality is essential for elucidating biomolecular interactions in pharmacology[7, 8] and biochemistry.[6, 9-11]

Various analytical techniques have been developed to investigate biomolecular chirality, including high-performance liquid chromatography,[12, 13] circular dichroism (CD) spectroscopy,[14-16] and surface-enhanced Raman optical activity.[17-19] Among these methods, CD spectroscopy is promising for its advantages of label-free,[20] in situ detection[21] and high enantioselectivity (g-factor).[22] However, a fundamental limitation of CD spectroscopy arises from the orders-of-magnitude size mismatch between chiral molecules — typically ranging from several to tens of nanometers) and circularly polarized light — hundreds of nanometers,[23, 24] which results in weak light–matter coupling and a characteristically low signal enhancement.[25] This low enhancement makes it challenging to detect chirality with dilute samples or probe complex sensing environments. Consequently, enhancing interactions between molecules and electric field interactions through tailored electromagnetic structures is gaining relevance in chiroptical sensing.

Extensive efforts have focused on achieving superchirality, where local optical chirality surpass that of circularly polarized light. [26-28] However, most existing designs typically contain a highly localized enhancement regions (hotspots) which are often too small to provide sufficient light matter interaction in certain cases. For example, when molecules are sparsely dispersed in solution, the probability of their overlapping with a hotspot is very low.[29, 30] For large chiral molecules (e.g. Amyloid proteins), the hotspot may interact with only a small portion of the molecule, resulting in limited enhancement.[31] To address these issues, the three-dimensional (3D) plasmonic design has been introduced to significantly increase the effective sensing volume and improve spatial coverage compared to conventional designs.[32, 33] However, expanding the mode volume often occurs at the expense of peak field intensity.

Here, we utilize 3D helical nanostructures to enhance chiroptical sensing performance at a large mode volume. We introduced a figure of merit (FOM) that incorporates the total integrated chiral field and the averaged chiral field. Because these two quantities exhibit opposite trends, the proposed FOM allows us to determine the optimal structure. The structures are designed for two common sensing scenarios in CD spectroscopy: (1) clusters of randomly oriented small molecules, and (2) a large chiral molecule, as schematically shown in Figure 1a. In both scenarios, effective CD sensing requires not only strong chiral field enhancement but also sufficient spatial accessibility of the enhanced field to the analytes. To meet these requirements, we implement a FOM-guided helix-geometry evolution that systematically tunes the key structural degrees of freedom to balance field enhancement and sensing volume. As shown in Figure 1b, the design is evolve by: (i) transforming a planar split-ring resonator (SRR) into a 3D helix to induce 3D chirality; (ii) winding escalation to expand the superchiral volume; (iii) promoting helical order from a single- to multi-strand helix configuration to enhance energy coupling; and (iv) transverse dilation to balance field enhancement and accessible sensing volume. This FOM-guided strategy provides wider superchiral field coverage and improved accessibility to chiral analytes, offering a practical framework for developing next-generation 3D CD sensing platforms.

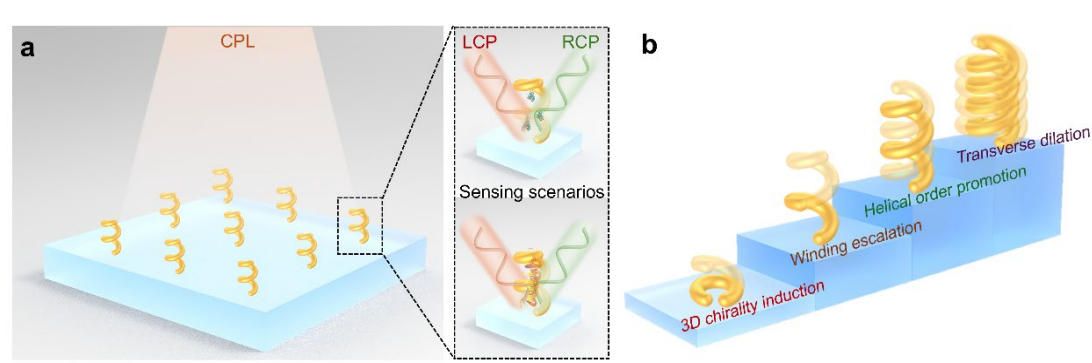

**Figure 1.** Conceptual framework of FOM-guided 3D chiroptical sensing. (a) Schematic illustration of CD chiral sensing using an array of helical nanostructures illuminated by circularly polarized light. The inset shows the two common sensing scenarios: clusters of randomly oriented small molecules (top) and a large chiral molecule (bottom). (b) The FOM-guided optimization of helix geometry. The evolution involves transforming a planar SRR to a 3D helix, winding escalation to expand the superchiral volume, promoting helical order from single- to multi-strand helix configuration, and transverse dilation.

Circular dichroism (CD) measures the differential absorption of left- and right-handed circularly polarized light (CPL) by chiral molecules. The chiral molecule absorption rate is described by:[28]

$$A^{\pm} = \frac{\omega}{2} Im(\mathbf{E}^* \cdot \mathbf{p} + \mathbf{B}^* \cdot \mathbf{m}) \tag{1}$$

where $A^+$ and $A^-$ represent the absorption rates for left- and right-handed CPL, respectively; $\mathbf{p}$ and $\mathbf{m}$ are the electric and magnetic dipole moments of the molecule; $\mathbf{E}$ and $\mathbf{B}$ are the complex vectors representing local electric and magnetic fields, respectively; stars here represent complex conjugation.

For a monochromatic CPL plane wave, the enantioselectivity ($g$) introduced by Tang and Cohen[28] is expressed as:

$$g = 2 \frac{A^+ - A^-}{(A^+ + A^-)} \tag{2}$$

The absorption rate difference can further be related to the optical chirality of the field as:[34]

$$A^+ - A^- = -\frac{2}{\varepsilon_0} Im(\xi)(C^+ - C^-) \tag{3}$$

where $\xi$ is the isotropic mixed electric-magnetic dipole polarizability, and $C^{\pm}$ are the optical chiral density values under left- and right-handed CPL illumination.

The interaction between an optical field and chiral molecules is determined by the optical chiral density ($C$), a quantity that characterizes the helicity of the electromagnetic field. This form of optical chirality was first introduced in 1964 by Lipkin[35] and serves as a key parameter for quantifying the chirality of light. The optical chiral density can be expressed as:

$$C = \frac{\varepsilon_0}{2} \mathbf{E} \cdot (\nabla \times \mathbf{E}) + \frac{1}{2\mu_0} \mathbf{B} \cdot (\nabla \times \mathbf{B}) = -\frac{\varepsilon_0 \omega}{2} Im(\mathbf{E}^* \cdot \mathbf{B}) \tag{4}$$

where $\varepsilon_0$ and $\mu_0$ are the vacuum permittivity and permeability, respectively; $\omega$ is the angular frequency.

The maximum chiral density achievable by a circularly polarized plane wave in the vacuum is given by:

$$C_{CPL} = \pm \frac{\varepsilon_0 \omega}{2c} E_0^2 \tag{5}$$

where $c$ is the speed of light, and signs + and − correspond to left- and right-handed CPL, respectively. Thus, the normalized optical chiral density ($\hat{C}$) introduced by nanostructures can be defined as:

$$\hat{C} = \frac{C}{|C_{CPL}|} = c \frac{Im(\mathbf{E}^* \cdot \mathbf{B})}{|E_0|^2} \tag{6}$$

Although $\hat{C}$ is useful for evaluating the strength of local chiral enhancement, it does not account for the volume where the chiral field is enhanced. This means, optimizing nanostructure with only $\hat{C}$ may result in the structure that provide very strong point-like hotspot, but rest region may have negligible $\hat{C}$. This tiny volume with enhanced $\hat{C}$ is problem when target sample includes sparse chiral molecules, or when chiral molecules are considerably larger than the mode volume. To address this problem, we propose a FOM that considers not only the chiral enhancement but also the spatial coverage of the superchiral field. The FOM is defined as:

$$FOM = C_{avg} \cdot C_{total} \tag{7}$$

where the total integrated superchiral field volume is:

$$C_{total} = \iiint_{V^\pm} \hat{C}^\pm(r) \, dV \tag{8}$$

and the average normalized chiral density is:

$$C_{avg} = \frac{1}{V^\pm} \iiint_{V^\pm} \hat{C}^\pm(r) \, dV \tag{9}$$

Here, the effective sensing volume ($V$) is defined as a region where $|\hat{C}| > 1$. The + and – signs correspond to left- and right-handed enhancement regions, respectively.

We designed the superchiral field by breaking the out-of-plane symmetry of a 2D split ring resonator (SRR). By progressively elevating one end of the SRR, it transformed into a 3D helix as shown in Figure 2a. The design parameters of the helix were selected in accordance with the feature-size constraints of two-photon lithography.[36, 37] Details of the simulation setting were provided in Supplementary Note 1. To evaluate the chiroptical sensing performance, the FOM values were simulated for helix pitch $p = 0 - 1000$ nm and wavelengths $\lambda = 3 - 5$ μm, as shown in Figure 2b. For every value of $p$, we extracted the peak FOM value across $\lambda$. As shown in Figure 2c, the peak FOM value generally increases up to $p \sim 400$ nm, beyond which the value decreases. This trend can be explained by separating the peak FOM values into their $C_{avg}$ and $C_{total}$ components. When the out-of-plane symmetry of the SRR was broken initially, the superchiral field around the structure was strongly enhanced, causing $C_{total}$ to increase with $p$. However, the value of $C_{total}$ reaches saturation after $p \sim 600$ nm. Consequently, $C_{avg}$ decreased due to dilution of the superchiral field, which can be observed in Figure 2d. This behavior highlighted that the FOM can effectively balance the enhancement strength and the spatial coverage of the superchiral field, thus improving chiroptical sensing performance.

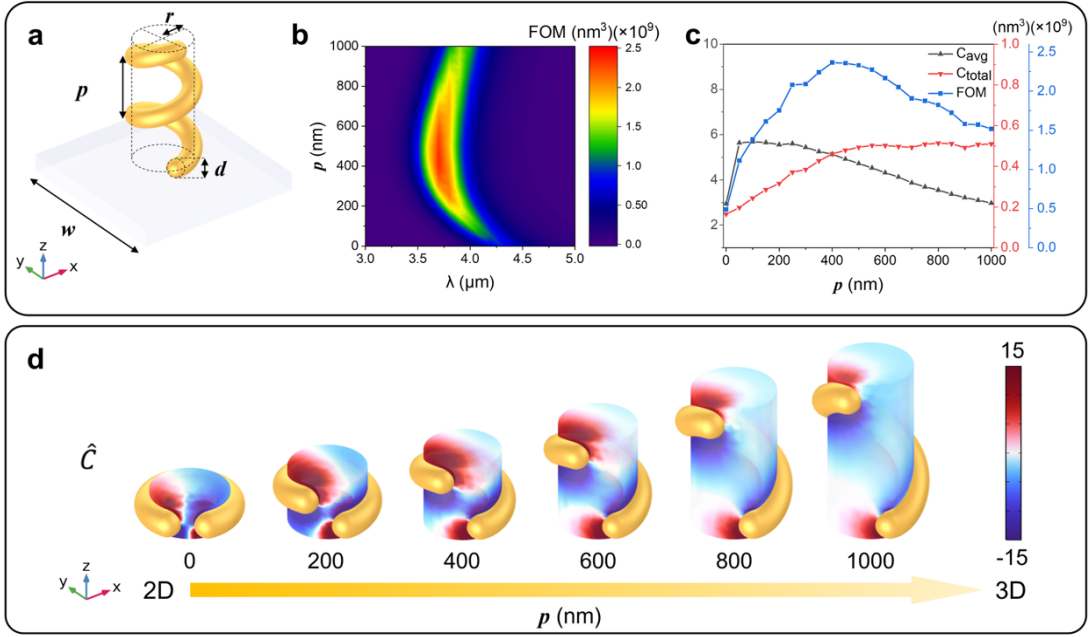

**Figure 2.** Evolution of superchiral field characteristics of single-strand helix driven by 3D chirality induction. (a) Schematic illustration of the helical geometry and defining parameters: periodic length $w$ = 1600 nm, helix radius $r$ = 250 nm, number of turns $\alpha$ = 0.83, wire diameter $d$ = 180 nm, and variable helix pitch $p$ from 0 to 1000 nm. (b) Simulated FOM (figure of merit) distribution of single-strand helix for varying $p$ and wavelength $\lambda$. (c) Graph of FOM, $C_{avg}$ (average normalized chiral density), $C_{total}$ (total integrated superchiral field volume), vs. $p$. (d) Evolution of normalized chiral density field $\hat{C}$ distribution by progressively augmenting the pitch.

To further enhance the performance of chiroptical sensing, we investigated the impact of the winding escalation of the single-strand helix, as illustrated in Figure 3a. A significant advantage of the helical geometry lies in shape self-similarity, which leads to two key benefits for chiroptical sensing: (i) preserving the overall normalized chiral density field distribution, and (ii) enhancing the FOM sensing performance due to enlarged effective sensing volume. The resonant modes exhibited similar normalized chiral density field distribution across different values of $\alpha$ (number of turns in the helix), as depicted in Figure 3b. The FOM distribution clearly illustrated the evolution of plasmonic resonant modes in the mid-infrared range, as shown in Figure 3c. The resonance originates from standing-wave nodes formed by electric dipole distributions along the helical path under circularly polarized light.[38] The node configuration changed as the helix is extended, resulting in the evolution of resonant modes. As shown in Figure 3d, the strongest resonant mode occurred at $\alpha$ = 5.67, confirming the improvement of chiroptical sensing performance during winding escalation.

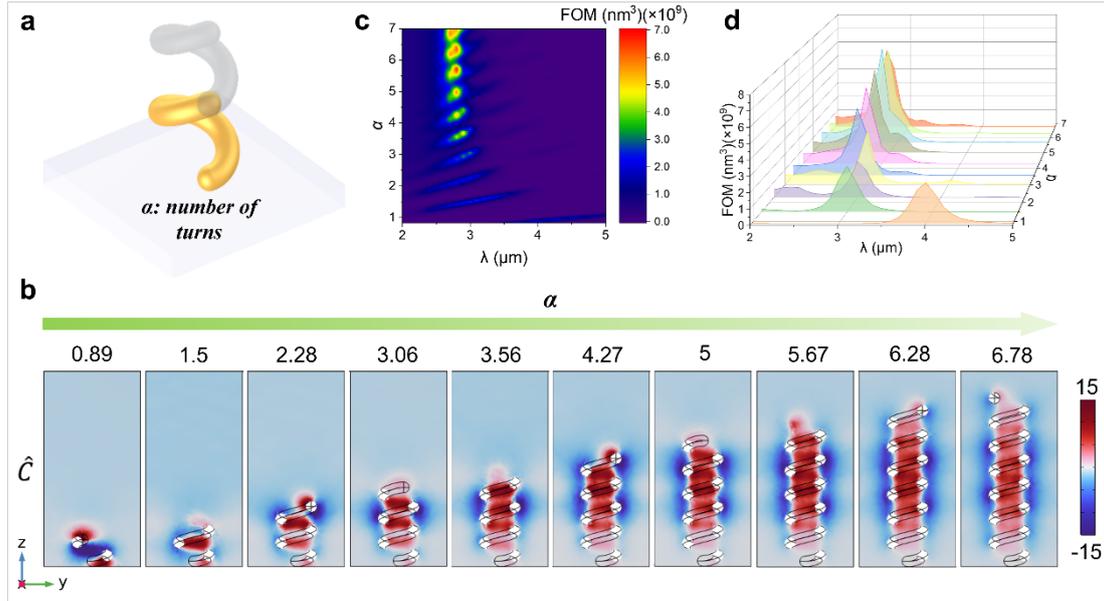

**Figure 3.** Scalable FOM enhancement enabled by winding escalation in a single-strand helix. The simulated structure is defined by the following parameters: $w$ = 1600 nm, $r$ = 250 nm, $d$ = 180 nm, $p$ = 400 nm, and variable $\alpha$ from 0.8 to 7. (a) Schematic illustration of the winding escalation process. (b) The preservation of normalized chiral density field distribution during winding escalation. (c) Simulated FOM distribution for varying $\alpha$ and wavelength $\lambda$. (d) Waterfall plot of the FOM vs. $\lambda$ for specific values of $\alpha$. The maximum FOM values occurred at $\alpha$ = 5.67.

We can further improve the FOM by promoting helix order, starting from a single-strand to a triple-strand helix, as illustrated in Figure 4a. The additional strands inherently required an increase in the helix pitch to accommodate multiple strands without geometric overlap. Based on this geometric constraint, our simulation demonstrated that the single-strand helix and multiple-strand helix had similar superchiral field distribution (see Supplementary Note 2). However, the multiple-strand helix exhibited a larger $C_{avg}$ than the single-strand helix due to the enhanced coupling efficiency and energy confinement (see Supplementary Note 3). To compare the chiroptical performance across different helical structures, we extracted each resonant mode of FOM spectra during the winding escalation for the single- and multiple-strand helix. As shown in Figure 4b, the peak FOM values of the multiple-strand helix were about an order of magnitude higher than the single-strand helix. We then plotted the peak FOM values vs. effective sensing volume for different helix orders. As shown in Figure 4c, helix order promotion increased the effective sensing volume, indicating that the superchiral fields were spatially expanded rather than confined to highly localized hotspots. This increase in sensing volume improves the accessibility to the chiral analyte, which is particularly beneficial for practical sensing scenarios involving clusters of randomly oriented small molecules or large chiral analytes. Hence, through helix order promotion, a multiple-strand helix provided greater chiroptical sensing performance and sensing volume.

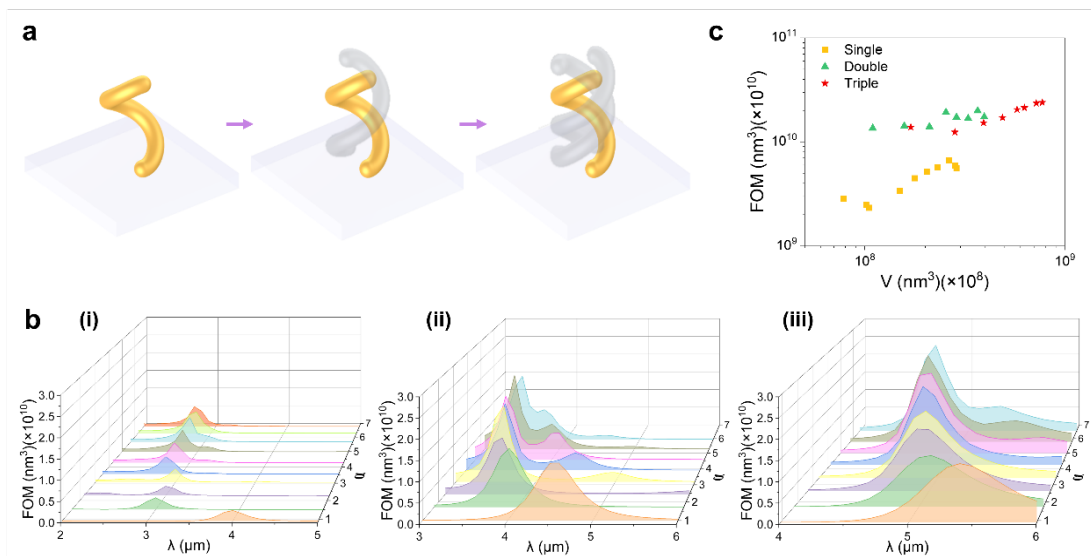

**Figure 4.** Enhancement of FOM and effective sensing volume $V$ through helix order promotion. (a) Schematic of single-strand, double-strand, and triple-strand helix. The structures are defined by parameters: $w = 1600$ nm, $r = 250$ nm, $d = 180$ nm, variable $\alpha$ from 0.8 to 7, and $p = 400/550/1000$ nm (for the single-strand/ double-strand/ triple-strand helix, respectively). (b) Waterfall plots of FOM vs. wavelength $\lambda$ for specific values of $\alpha$ in (i) single-strand helix, (ii) double-strand helix, and (iii) triple-strand helix. (c) Relationship between FOM and $V$ for different helix orders.

Lastly, we investigated the transverse dilation effect of a triple-strand helix, as illustrated in Figure 5a. We simulated the FOM distribution for varying helix radius ($r = 200$nm – 600nm) and wavelength ($\lambda = 3 – 10$ μm). As shown in Figure 5b, the FOM distribution exhibited a distinct resonance trajectory that redshifts with increasing radius. For every value of $r$, we extracted the peak FOM value and its $C_{avg}$ and $C_{total}$ components across $\lambda$. As shown in Figure 5c, an excessive increase in $r$ leads to a decrease in $C_{avg}$, $C_{total}$, and the peak FOM value. This trend can be explained by weaker intensity due to the expansion of the superchiral field region, as shown in Figure 5d. These results reveal a fundamental trade-off between chiral enhancement and spatial coverage, underscoring the need for FOM-based optimization in high-performance chiroptical sensing.

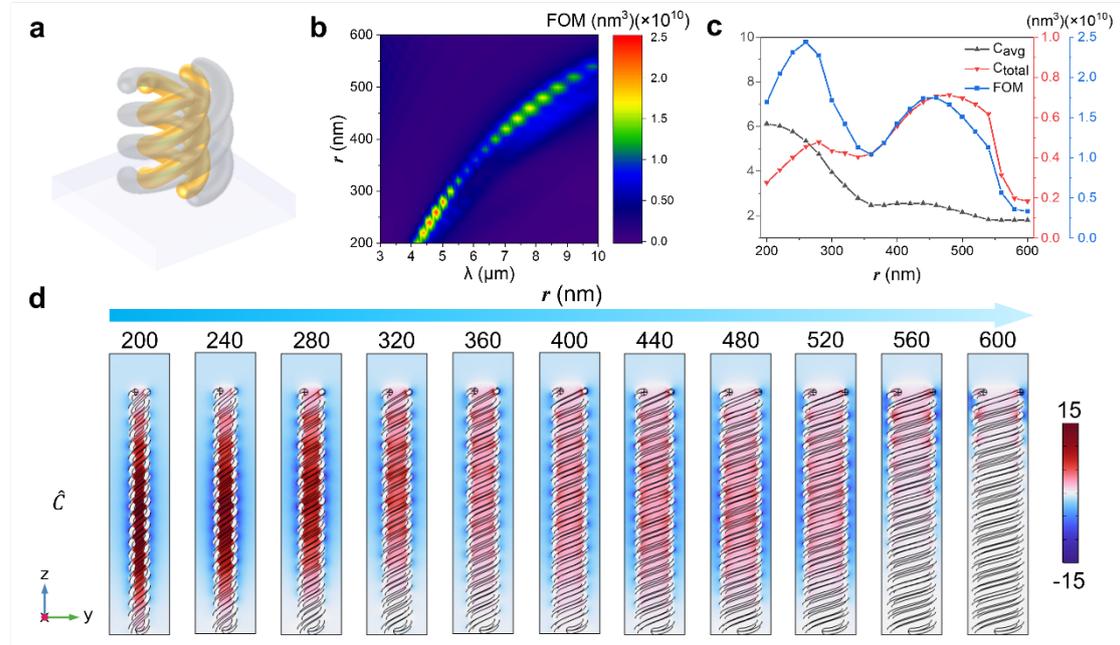

**Figure 5.** Effect of transverse dilation on superchiral field distribution and FOM. The simulated structure is defined by the parameters: $w = 1600$ nm, $p = 1000$ nm, $d = 180$ nm, $\alpha = 6.55$, and variable $r$ from 200 to 600 nm. (a) Schematic illustration depicts the process of transverse dilation. (b) Simulated FOM distributions exhibiting a redshifted resonance trajectory for varying $p$ and wavelength $\lambda$. (c) Graph of FOM, $C_{avg}$ (average normalized chiral density), $C_{total}$ (total integrated superchiral field volume), vs. $r$. (d) Normalized chiral density distribution during transverse dilation.

Our helix-geometry evolution strategy can enlarge the sensing volume without compromising the distribution or enhancement strength of fields. The triple-strand helix design achieved a FOM value of $2.43 \times 10^{10}$ nm$^3$, significantly outperforming conventional chiral nanostructures,[33, 39] as summarized in Table 1. This performance reflected the strong chiroptical response enabled by geometry evolution that maximizes both sensing volume and chiral signal enhancement.

**Table 1.** FOM values between our design and conventional chiroptical sensing design.

| Type | Structure | V (nm³) | FOM (nm³) |
|---|---|---|---|
| 2D | Gammadion[39] | $2.02\times10^6$ | $3.37\times10^7$ |
| 2D | Double spiral[39] | $1.64\times10^6$ | $5.57\times10^8$ |
| 3D | Single helix[33] | $3.61\times10^7$ | $3.53\times10^8$ |
| 3D | Four-helix[33] | $8.83\times10^8$ | $5.54\times10^9$ |
| 3D | Triple-strand helix (this work) | $3.44\times10^9$ | $2.43\times10^{10}$ |

To verify the relationship between the FOM and enhancement of molecular CD signals in common sensing scenarios, we conducted full wave simulation by embedding chiral material at the core of the helical structures, as illustrated in Figure 6a. The chiral analyte property was defined by wavelength-independent Pasteur parameters (see Supplementary Note 4).[40, 41] The simulated CD spectra for the triple-strand helix are shown in Figure 6b. The maximum CD signals occurred at the resonance wavelengths of the bare helical structures, confirming that the observed CD enhancement originates from the superchiral fields supported by the structures (see Supplementary Note 5). We analyzed how the maximum CD signal ($CD_{max}$) varies with different values of FOM in Figure 6c, which showed a strong linear relationship between FOM and CD signal with a correlation coefficient of 0.9256, confirmed by applying the analysis of variance (ANOVA) method at a 95% confidence level (see Supplementary Note 6). Hence, the proposed FOM metric was reliable for evaluating CD enhancement performance.

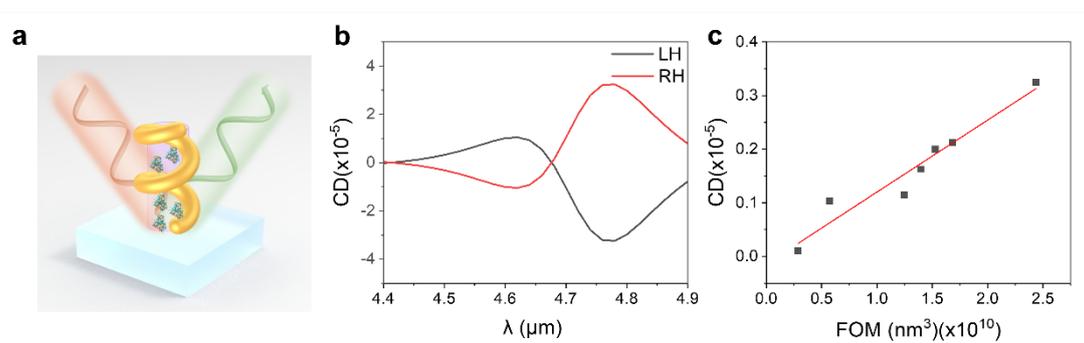

**Figure 6** Simulation of the chiroptical response of chiral analytes situated within the superchiral fields of a 3D helix. (a) Schematic of the circular dichroism sensing scheme. (b) The background-subtracted CD signal spectrum for triple-strand helix structures. The results are shown for a structure with the parameters: $w$ = 1600 nm, $d$ = 180 nm, $p$ = 1000 nm, $\alpha$ = 6.55, and $r$ = 260 nm. (c) The variations of CD$_{max}$ intensity with different values of FOM. The red line represents the linear regression.

In conclusion, the newly introduced FOM for chiroptical sensing performance was applied to identify optimal helical nanostructures that exhibit: (1) a substantial increase in chiroptical sensing performance, and (2) an expanded superchiral field region more accessible to chiral analytes. This design approach is distinct from previous works that only optimize chiral enhancement. A strong linear correlation is found between the CD signal and the FOM, validating it as a robust metric for evaluating superchiral field configurations. Notably, the triple-helix structure realizes a record-high FOM of $2.43 \times 10^{10}$ nm$^3$, outperforming previously reported 2D and 3D architectures. This design framework opens avenues for applications such as tailoring chiral sensors that operate in complex biological environments or targeting specific biomolecular assemblies with varying spatial distributions. Future works can explore dynamic or reconfigurable chiroptical platforms for spectrally tunable responses and integration with microfluidic platforms for real-time biosensing. In addition, this strategy can be extended to near-IR or visible spectrums to broaden its utility across diverse molecular detection schemes.

## Author Information

**Corresponding Authors**

Chia-Ming Yang - Department of Electronic Engineering, Chang Gung University, Taoyuan City 33302, Taiwan (R.O.C.); Institute of Electro-Optical Engineering, Chang Gung University, Taoyuan City 33302, Taiwan (R.O.C.); Department of Neurosurgery, Chang Gung Memorial Hospital at Linkou, Taoyuan City 33305, Taiwan (R.O.C.); Department of Materials Engineering, Ming Chi University of Technology, New Taipei City 243303, Taiwan (R.O.C.); Department of Electronic Engineering, Ming Chi University of Technology, New Taipei City 243303, Taiwan (R.O.C.)

Sejeong Kim - Department of Electrical and Computer Engineering, Sungkyunkwan University (SKKU), Suwon 16419, Republic of Korea

Hongtao Wang - Engineering Product Development, Singapore University of Technology and Design, Singapore 487372, Singapore

Joel K.W. Yang - Engineering Product Development, Singapore University of Technology and Design, Singapore 487372, Singapore; Singapore-HUJ Alliance for Research and Enterprise (SHARE), The Smart Grippers for Soft Robotics (SGSR) Programme, Campus for Research Excellence and Technological Enterprise (CREATE), Singapore 138602, Singapore

**Authors**

Chia-Te Chang - Engineering Product Development, Singapore University of Technology and Design, Singapore 487372, Singapore

Xiaoyan Zhou - Engineering Product Development, Singapore University of Technology and Design, Singapore 487372, Singapore

Dmitrii Gromyko - Science, Mathematics, and Technology (SMT), Singapore University of Technology and Design, Singapore 487372, Singapore; Department of Electrical and Computer Engineering, National University of Singapore, Singapore 117583, Singapore

John You En Chan - Engineering Product Development, Singapore University of Technology and Design, Singapore 487372, Singapore

Lin Wu - Science, Mathematics, and Technology (SMT), Singapore University of Technology and Design, Singapore 487372, Singapore

**Author Contributions**

[†]C.-T.C. and X.Z. contributed equally to this work.

**Notes**

The authors declare no competing financial interest.

# Acknowledgments

J.K.W.Y. acknowledges funding support from the National Research Foundation (NRF) Singapore (NRF-CRP29-2022-0001), the NRF Investigatorship Award (NRF-NRFI06-2020-0005) and National Research Foundation, Prime Minister's Office, Singapore under Campus of Research Excellence and Technological Enterprise (CREATE) programme. C.-M.Y. acknowledges funding support Chang Gung University, Taiwan with contract no. of UERPD2Q0051.

## References

(1) Brookes, J. C.; Horsfield, A.; Stoneham, A. Odour character differences for enantiomers correlate with molecular flexibility. *Journal of the Royal Society Interface* **2009**, *6* (30), 75-86.

(2) Islam, M. R.; Mahdi, J. G.; Bowen, I. D. Pharmacological importance of stereochemical resolution of enantiomeric drugs. *Drug Safety* **1997**, *17* (3), 149-165.

(3) Smith, S. W. Chiral toxicology: it's the same thing… only different. *Toxicological Sciences* **2009**, *110* (1), 4-30.

(4) Liu, W.; Tang, M. Enantioselective activity and toxicity of chiral herbicides. *Herbicides-Mechanisms and Mode of Action* **2011**, 63-80.

(5) Ding, F.; Peng, W.; Peng, Y.-K.; Liu, B.-Q. Estimating the potential toxicity of chiral diclofop-methyl: Mechanistic insight into the enantioselective behavior. *Toxicology* **2020**, 152446.

(6) Törnquist, M.; Linse, S. Chiral selectivity of secondary nucleation in amyloid fibril propagation. *Angewandte Chemie* **2021**, *60*, 24008-24011.

(7) Singh, J.; Hagen, T. Chirality and biological activity. *Burger's Medicinal Chemistry and Drug Discovery* **2010**, 127-166.

(8) Maher, T.; Johnson, D. Review of chirality and its importance in pharmacology. *Drug Development Research* **1991**, *24*.

(9) Gal, J. Molecular chirality: language, history, and significance. *Topics in Current Chemistry* **2015**, *340*, 1-20.

(10) Malloy, J. F.; Millsaps, C.; Narasimhan, K.; Slocombe, L.; Mathis, C.; Cronin, L.; Walker, S. I. The emergence of chirality from metabolism. *arXiv preprint arXiv:2505.01056* **2025**.

(11) Barron, L. Chirality and life. *Space Science Reviews* **2008**, *135*, 187-201.

(12) Bertucci, C.; Tedesco, D. Advantages of electronic circular dichroism detection for the stereochemical analysis and characterization of drugs and natural products by liquid chromatography. *Journal of Chromatography. A* **2012**, *1269*, 69-81.

(13) Siegle, A.; Trapp, O. Hyphenation of hadamard encoded multiplexing liquid chromatography and circular dichroism detection to improve the signal-to-noise ratio in chiral analysis. *Analytical Chemistry* **2015**, *87 23*, 11932-11934.

(14) Molteni, E.; Onida, G.; Ceccarelli, M.; Cappellini, G. Ab initio spectroscopic investigation of pharmacologically relevant chiral molecules: the cases of avibactam, cephems, and idelalisib as benchmarks for antibiotics and anticancer drugs. *Symmetry* **2021**, *13*, 601.

(15) Taniguchi, T.; Usuki, T. Circular dichroism spectroscopy. *Supramolecular Chemistry: From Molecules to Nanomaterials* **2012**.

(16) Lassen, P. R. Structural characterization of chiral molecules using vibrational

circular dichroism spectroscopy. **2006**.

(17) Lee, S. G.; Kwak, S.; Son, W.; Kim, S.; Nam, K.; Lee, H.-Y.; Jeong, D. Chiral-induced surface-enhanced raman optical activity on a single-particle substrate. *Analytical Chemistry* **2024**.

(18) Das, M.; Gangopadhyay, D.; Andrushchenko, V.; Kapitán, J.; Bouř, P. Bisignate surface-enhanced raman optical activity with analyte-capped colloids. *ACS Nano* **2025**, *19*, 10412-10420.

(19) Li, H.; Zhang, J.; Jiang, L.; Yuan, R.; Yang, X. Chiral plasmonic Au-Ag core shell nanobipyramid for SERS enantiomeric discrimination of biologically relevant small molecules. *Analytica Chimica Acta* **2023**, *1239*, 340740.

(20) Farahani, N.; Behmanesh, M.; Ranjbar, B. Evaluation of rationally designed label-free stem-loop DNA probe opening in the presence of miR-21 by circular dichroism and fluorescence techniques. *Scientific Reports* **2020**, *10*.

(21) Fiedler, S.; Cole, L.; Keller, S. Automated circular dichroism spectroscopy for medium-throughput quantification of protein conformation. *Biophysical Journal* **2013**, *104*.

(22) Reetz, M. T.; Kühling, K. M.; Hinrichs, H.; Deege, A. Circular dichroism as a detection method in the screening of enantioselective catalysts. *Chirality: The Pharmacological, Biological, and Chemical Consequences of Molecular Asymmetry* **2000**, *12* (5-6), 479-482.

(23) Hu, J.; Xiao, Y.; Zhou, L.; Jiang, X.; Qiu, W.; Fei, W.; Chen, Y.; Zhan, Q. Ultra-narrow-band circular dichroism by surface lattice resonances in an asymmetric dimer-on-mirror metasurface. *Optics Express* **2022**, *30 10*, 16020-16030.

(24) Lorin, M.; Delépée, R.; Maurizot, J.; Ribet, J.; Morin, P. Sensitivity improvement of circular dichroism detection in HPLC by using a low-pass electronic noise filter: Application to the enantiomeric determination purity of a basic drug. *Chirality* **2007**, *19 2*, 106-113.

(25) Covington, C.; Polavarapu, P. Similarity in dissymmetry factor spectra: a quantitative measure of comparison between experimental and predicted vibrational circular dichroism. *The Journal of Physical Chemistry. A* **2013**, *117 16*, 3377-3386.

(26) Hendry, E.; Carpy, T.; Johnston, J.; Popland, M.; Mikhaylovskiy, R. V.; Lapthorn, A. J.; Kelly, S. M.; Barron, L. D.; Gadegaard, N.; Kadodwala, M. Ultrasensitive detection and characterization of biomolecules using superchiral fields. *Nature Nanotechnology* **2010**, *5* (11), 783-787.

(27) Xu, C.; Ren, Z.; Zhou, H.; Zhou, J.; Ho, C. P.; Wang, N.; Lee, C. Expanding chiral metamaterials for retrieving fingerprints via vibrational circular dichroism. *Light: Science & Applications* **2023**, *12* (1), 154.

(28) Tang, Y.; Cohen, A. E. Optical chirality and its interaction with matter. *Physical*

*Review Letters* **2010**, *104* (16), 163901.

(29) Powis, I. Photoelectron circular dichroism of the randomly oriented chiral molecules glyceraldehyde and lactic acid. *The Journal of Chemical Physics* **2000**, *112* (1), 301-310.

(30) Beaulieu, S.; Comby, A.; Descamps, D.; Fabre, B.; Garcia, G. A.; Géneaux, R.; Harvey, A. G.; Légaré, F.; Mašín, Z.; Nahon, L.; et al. Photoexcitation circular dichroism in chiral molecules. *Nature Physics* **2018**, *14* (5), 484-489.

(31) Rodger, A.; Pančoška, P. Circular dichroism in analysis of biomolecules. In *Encyclopedia of Analytical Chemistry*, pp 1-41.

(32) Tseng, M. L.; Lin, Z.-H.; Kuo, H. Y.; Huang, T.-T.; Huang, Y.-T.; Chung, T. L.; Chu, C. H.; Huang, J.-S.; Tsai, D. P. Stress-induced 3D chiral fractal metasurface for enhanced and stabilized broadband near-field optical chirality. *Advanced Optical Materials* **2019**, *7* (15), 1900617.

(33) Schäferling, M.; Yin, X.; Engheta, N.; Giessen, H. Helical Plasmonic Nanostructures as Prototypical Chiral Near-Field Sources. *ACS Photonics* **2014**, *1* (6), 530-537.

(34) Schäferling, M. *Chiral nanophotonics*; Springer, 2017.

(35) Lipkin, D. M. Existence of a new conservation law in electromagnetic theory. *Journal of Mathematical Physics* **1964**, *5* (5), 696-700.

(36) Zhou, X.; Wang, H.; Liu, S.; Wang, H.; Chan, J. Y. E.; Pan, C.-F.; Zhao, D.; Yang, J. K. W.; Qiu, C.-W. Arbitrary engineering of spatial caustics with 3D-printed metasurfaces. *Nature Communications* **2024**, *15* (1), 3719.

(37) Zhang, W.; Min, J.; Wang, H.; Wang, H.; Li, X. L.; Ha, S. T.; Zhang, B.; Pan, C.-F.; Li, H.; Liu, H.; et al. Printing of 3D photonic crystals in titania with complete bandgap across the visible spectrum. *Nature Nanotechnology* **2024**, *19* (12), 1813-1820.

(38) Höflich, K.; Feichtner, T.; Hansjürgen, E.; Haverkamp, C.; Kollmann, H.; Lienau, C.; Silies, M. Resonant behavior of a single plasmonic helix. *Optica* **2019**, *6* (9), 1098-1105.

(39) Schäferling, M.; Dregely, D.; Hentschel, M.; Giessen, H. Tailoring enhanced optical chirality: design principles for chiral plasmonic nanostructures. *Physical Review X* **2012**, *2* (3), 031010.

(40) Nesterov, M. L.; Yin, X.; Schäferling, M.; Giessen, H.; Weiss, T. The role of plasmon-generated near fields for enhanced circular dichroism spectroscopy. *ACS Photonics* **2016**, *3* (4), 578-583.

(41) Solomon, M. L.; Abendroth, J. M.; Poulikakos, L. V.; Hu, J.; Dionne, J. A. Fluorescence-detected circular dichroism of a chiral molecular monolayer with dielectric metasurfaces. *Journal of the American Chemical Society* **2020**, *142* (43), 18304-18309.